\documentclass[journal=jcisd8,manuscript=article]{achemso}
\setkeys{acs}{etalmode=truncate,maxauthors=0}

\usepackage[version=3]{mhchem} %
\usepackage{amssymb}
\usepackage{amsfonts}
\usepackage{amsmath}
\usepackage{multirow}
\usepackage{booktabs}
\usepackage{epsfig}
\usepackage{pdfpages}
\usepackage{epstopdf}
\usepackage{lscape}
\usepackage{rotating}
\usepackage{graphicx}
\usepackage{placeins}
\usepackage{subfigure}
\usepackage{caption}
\usepackage{hyperref}
\usepackage{soul}

\usepackage{xr}
\makeatletter

\newcommand*{\addFileDependency}[1]{%
\typeout{(#1)}%
\@addtofilelist{#1}
\IfFileExists{#1}{}{\typeout{No file #1.}}
}\makeatother

\newcommand*{\myexternaldocument}[1]{%
\externaldocument{#1}%
\addFileDependency{#1.tex}%
\addFileDependency{#1.aux}%
}
\myexternaldocument{SI}

\newcommand{\pK}             {\ensuremath{\mathrm{p}K}}
\newcommand{\pKa}            {\ensuremath{\mathrm{p}K_\mathrm{a}}}
\newcommand{\pKatitle}       {\boldmath$\mathrm{p}K_{\mathrm{a}}$}

\newcommand{\pKmod}          {\ensuremath{\mathrm{p}K^{\mathrm{mod}}}}

\makeatletter
\def\acs@author@fnsymbol#1{}
\makeatother

\author{Tom\'as F. D. Silva$^*$}
\affiliation{Scuola Internazionale Superiore di Studi Avanzati, 34136 Trieste, Italy}

\author{Giovanni Bussi}
\affiliation{Scuola Internazionale Superiore di Studi Avanzati, 34136 Trieste, Italy}

\email{tfernand@sissa.it}

\title{Characterizing RNA oligomers using Stochastic Titration Constant-pH Metadynamics simulations}
\begin{document}

\begin{abstract}
RNA molecules exhibit various biological functions intrinsically dependent on their diverse ecosystem of highly flexible structures. This flexibility arises from complex hydrogen-bonding networks defined by canonical and non-canonical base pairs that require protonation events to stabilize or perturb these interactions. Constant pH molecular dynamics (CpHMD) methods provide a reliable framework to explore the conformational and protonation space of dynamic structures and for robust calculations of pH-dependent properties, such as the \pKa{
} of titrable sites. Despite growing biological evidence concerning pH regulation of certain motifs and in biotechnological applications, pH-sensitive \textit{in silico} methods have rarely been applied to nucleic acids.
In this work, we extended the stochastic titration CpHMD method to include RNA parameters from the standard $\chi$OL3 AMBER force field and highlighted its capability to depict titration events of nucleotides in single-stranded RNAs.  
We validated the method using trimers and pentamers with a single central titrable site while integrating a well-tempered metadynamics approach into the st-CpHMD methodology (CpH-MetaD) using PLUMED. This approach enhanced the convergence of the conformational landscape and enabled more efficient sampling of protonation-conformation coupling. Our \pKa{} estimates agree with experimental data, validating the method's ability to reproduce electrostatic changes around a titrable nucleobase in single-stranded RNA. These findings provided molecular insight into intramolecular phenomena, such as nucleobase stacking and phosphate interactions, that dictate the experimentally observed \pKa{} shifts between different strands. Overall, this work validates both the st-CpHMD and the metadynamics integration as reliable tools for studying biologically relevant RNA systems. 
\end{abstract}

\section{Introduction}
\label{sec:intro}
 
pH is a ubiquitous environmental factor that significantly influences various biomolecules' structure, chemistry, and function. By influencing the protonation states of chemical moieties depending on their intrinsic \pKa{}, pH will affect their overall charge and thermodynamic equilibria~\cite{putnam2012,aoi2014}. This pH-dependent modulation is pivotal for regulating multiple biological processes, such as protein-protein interactions, nucleic acid binding, drug interactions, and conformational changes in response to shifts in the pH environment or nearby electrostatics~\cite{chen2015,wang2017}.
Usually, nucleic acids are not particularly sensitive to small physiological pH changes because the \pKa{} of nucleotides is quite far from the physiological one and even further away in common canonically base paired nucleotides~\cite{legault1997}. However, non-canonical base pairings and diverse arrays of inter- and intramolecular interactions promote more complex and pH-sensitive electrostatic environments. For instance, some non-canonical base pairs, such as the A$^+$-C wobble and G-C$^+$ Hoogsteen pairs, involve protonated forms of adenine (\pKa{}~3.5 - N1) and cytosine (\pKa{}~4.2 - N3)~\cite{jones2022}. In other instances, certain modified nucleobases, like 1-methyladenosine (\pKa{}~8.3 - N6) and 3-methylcytidine (\pKa{}~8.7 - N4), deprotonate at more basic conditions when compared to their unmodified counterparts, as they have significantly shifted down \pKa{} values closer to physiological pH (7.0-7.4)~\cite{jones2022}. These \pKa{} shifts highlight how the modified nucleobases are more prone to protonation and deprotonation events, which can energetically promote or hinder pH-dependent conformational rearrangements depending on the electrostatic environment and the medium pH.
Although several experimental studies have explored the \pKa{} and related properties of nucleobases~\cite{izatt1971,acharya2003}, nucleosides~\cite{clauwaert1968,gonzalez2015,gonzalez2018}, and nucleotides~\cite{acharya2003,acharya2004,acharya2005}, recent research has increasingly focused on the role of pH-dependent secondary and tertiary structures, such as i-motifs~\cite{collin1998,gueron2000} and triplexes~\cite{rhee1999,hu2017}. These structures are not only important for biological functions like catalysis and structural stability but also hold significant potential for biotechnological applications, including biosensors~\cite{mariottini2021,farag2021,cecconello2022}, drug delivery systems~\cite{miao2020}, and molecular switches~\cite{idili2014,mariottini2019}.
Yet detailed insights into the mechanisms of action of these systems are difficult to obtain through experimental protocols alone.

Molecular dynamics (MD) methods strongly complement experimental RNA studies by providing an atomistic description, although at shorter timescales~\cite{sponer2018}. 
However, standard MD simulations typically assume fixed protonation states based on the molecule's \pKa{} at physiological pH, ignoring the dynamic nature of protonation events modulated by its instantaneous 3D conformation and surrounding electrostatic environment. These limitations provide an incomplete picture of biomolecular behavior, especially in rich biological systems where protonation events drive relevant conformational interactions.
Constant-pH molecular dynamics (CpHMD) techniques have been developed to overcome these limitations. These methods introduce residue titration within an MD framework, hence enabling the prediction of protonation states and \pKa{} values of titratable groups. Importantly, it allows us to probe the dependency of the protonation state on the molecular conformation. CpHMD methods can be broadly categorized into continuous and discrete approaches. Continuous methods, typically based on $\lambda$ dynamics~\cite{kong1996} such as PHMD~\cite{lee2004,goh2012,goh2014} or PME-based CpHMD~\cite{huang2016} or the GROMACS scalable CpHMD version~\cite{aho2022}, sample both conformations and fractional protonation states by extending the Hamiltonian with a pH-dependent $\lambda$ particle with fictitious mass. These methods can be further divided based on implicit solvent models~\cite{lee2004}) or explicit solvent models~\cite{donnini2011,wallace2012,goh2014,aho2022}, and all have been successfully applied to different biomolecules (e.g., nucleic acids~\cite{goh2012,goh2013} and proteins~\cite{khandogin2005,lee2004})
using various force fields (e.g., CHARMM, AMBER).
Previous work on nucleic acid titration was done by Goh and coworkers~\cite{goh2012,goh2013}. This work focused on their continuous multi-site-$\lambda$-dynamics~\cite{knight2011} constant pH implementation (CPHMD$^{MS\lambda{}D}$) using the CHARMM force field. Adenosine and cytidine were used as model compounds in their \pKa{} calibration while obtaining good experimental agreement for their test compounds: adenosine monophosphate (AMP), cytidine monophosphate (CMP), and dinucleotides combinations (CYT-CYT, ADE-ADE, CYT-ADE). 
Discrete methods, on the other hand, usually employ a start-stop Monte Carlo(MC)/MD approach that accepts or rejects a protonation state switch, for each titrable residue, through a Metropolis criterion. The criterion depends on the protonation free-energies of a given residue, which is calculated on a frozen conformation using an implicit solvent model, either Generalized-Born 
as in PHREM~\cite{itoh2011} or Poisson-Boltzmann (PB)~\cite{baptista2002,burgi2002,dlugoz2004a} methods. For example, the stochastic titration constant-pH method (st-CpHMD) was originally developed by Baptista~\cite{baptista2002} for proteins~\cite{machuqueiro2006,machuqueiro2008,teixeira2016} using the GROMOS and, currently, also the CHARMM36 force field~\cite{sequeira2022}.  Within this approach, a model compound is used as a nonphysical fragment (i.e. the nucleobase) that encapsulates the \pKa{} of the chemical moiety within the molecule~\cite{machuqueiro2011}. For instance, the calibration of this \pKa{} (\pKmod{}) would be necessary for all nucleobases using experimental data of nucleosides~\cite{izatt1971}. With this procedure, the \pKmod{} is fine-tuned for any systematic deviations from experimental \pKa{} due to the PB parameters. The intricate details of the st-CpHMD method are extensively discussed in the literature~\cite{baptista2002,teixeira2005,machuqueiro2006,machuqueiro2011}. To the best of our knowledge, these discrete methods were never tested in nucleic acids.

Building on the previous work and standard aminoacid \pKa{} calibration protocol~\cite{machuqueiro2011}, we aim to extend the stochastic titration constant pH molecular dynamics (st-CpHMD) method to nucleic acids by adapting $\chi$OL3 AMBER force field parameters. Our approach entails parametrizing the charged states of non-modified RNA nucleobases through a RESP protocol, then a first-stage calibration of individual \pKmod{}'s using nucleoside data, and then a second-stage validation and recalibration of the \pKa{} values in oligonucleotides of varying sizes according to available experimental data~\cite{acharya2004,gonzalez2015,gonzalez2018}. This protocol integrates the effects of the phosphate backbone on \pKa{} shifts, analogous to aminoacid \pKmod{} calibration with pentapeptides~\cite{machuqueiro2011}, resulting in \pKa{} values more representative of biomolecular environments. Trimer and pentamer systems of each nucleobase flanked by non-titrating residues were simulated to assess the effects on the \pKa{} of increasing the backbone length by measuring their $\Delta$\pKa{}. %

Another focus of this work is to address the challenge posed by short flexible nucleotides in MD simulations~\cite{kuhrova2013,haldar2015,mlynsky2022} by integrating the CpHMD protocol with metadynamics~\cite{bussi2020} coupling to the PLUMED plugin.
This integration focuses on improving the sampling of system-specific collective variables (CVs) without introducing a bias in the protonation space of titratable sites. By the sites' conformation and protonation states being intrinsically coupled, enhanced conformational sampling improves the accuracy of average protonation and \pKa{} predictions, particularly for well-solvated sites. Our work presents a reliable and robust framework for the pH-dependent study of conformational dynamics in nucleic acids through a CpH-metaD approach.

\section{Methods}
\label{sec:methods}
\subsection{CpHMD simulation using AMBER $\chi$OL3 force field}

The CpHMD extension introduced in this paper builds upon the standard AMBER parameterization~\cite{cornell1995,perez2007,zgarbova2011} by solely introducing the charged states of the nucleotides using the neutral states as references. Other parameters were adapted from the original force field. Parameters were derived to be compatible with
the OPC water model \cite{izadi2014}. 

The new charge set parametrization used two steps of the restrained electrostatic potential (RESP) procedure~\cite{bayly1993} on optimized geometries of the protonated and deprotonated states of all nucleobases, with riboses replaced by methyl groups as in previous work~\cite{aduri2007}. The geometry optimization used the B3LYP functional with the 6-31G* dataset using the Gaussian16~\cite{g16} software. Then, RESP charges for each nucleobase's neutral and charged states were derived. The partial charges for the charged states were determined as follows: 1) we calculated, for each atom, the RESP partial charge difference between their charged and neutral states; and 2) we added the calculated RESP difference for each atom to the neutral $\chi$OL3 partial charges. The original charge set for the neutral states was preserved. Charges are shown in the Tables~\ref{Table-SI:charge_set_A}-~\ref{Table-SI:charge_set_G} of Supporting Information.

For the Poisson-Boltzmann calculations, we built the DelPhi databases for the atoms' radii and charges. In the radii procedure, Lennard-Jones parameters of all atom types were used, based on Lorentz-Berthelot combination rules~\cite{lorentz1881,berthelot1898}, to determine the radius of each atom against OPC water molecules, as done in other similar protocols~\cite{teixeira2005}. The atomic partial charges database was built from the original $\chi$OL3 force field partial charges, and the newly derived charge set for the charged states of the nucleobases.

Restricting the nucleobase net charge onto the model compound required tweaking the point charges on the C1', H1', and N9/N1 atoms (purines and pyrimidines, respectively) to ensure the nucleobase moiety had an integral charge.
These small modifications were also applied to the neutral state.
To validate them, we
performed 200~ns MD simulations for the canonical RNA nucleosides using the original $\chi$OL3 force field. Then we recomputed the energy of the same trajectories using topologies generated with our modified charge set. A comparative analysis was done using experimental NMR and computational $^3$J coupling data. All the mentioned parameters are found on the following GitHub repository: \url{https://github.com/Tomfersil/CpH-MetaD}.

\subsection{Systems setup}
\begin{figure}[H]
\centering
    \centering
    \includegraphics[width=1.0\textwidth]{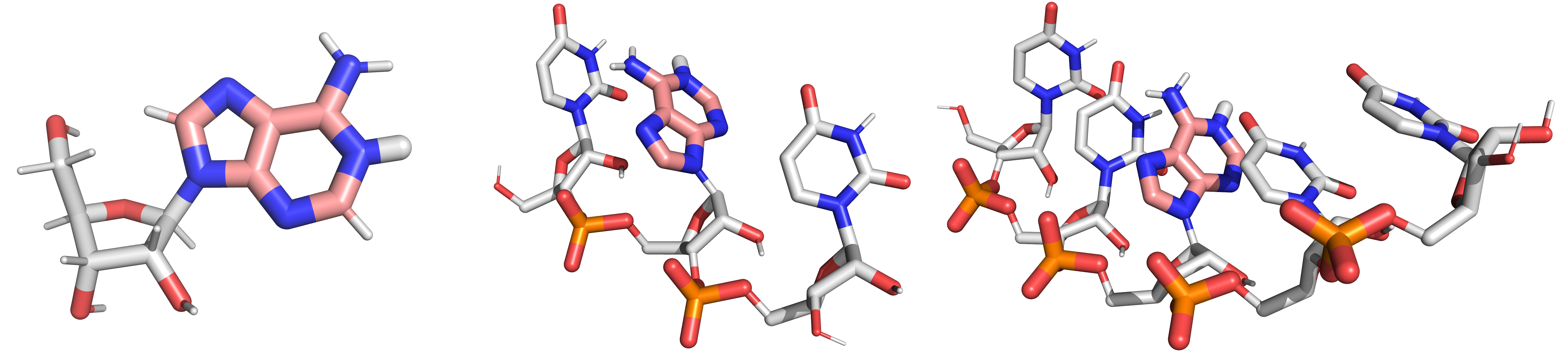}
\caption{Cartoon representation of the nucleoside, trimer, and pentamer Adenine systems simulated in this work. In the single-strand systems, non-titrable nucleotides are represented with white carbon and hydrogen atoms, and the phosphate group is shown in orange and red for the phosphorous and oxygen atoms, respectively. In every system, the titrable nucleobases and protons are represented in pink carbons, blue nitrogen, and thicker white hydrogen sticks. Each system is capped with hydroxyl groups at the 3$^{\prime}$ and 5$^{\prime}$ ends.}
\label{Fig:systems1}
\end{figure}

\begin{figure}[H]
\centering
\begin{subfigure}
    \centering
    \includegraphics[width=0.45\textwidth]{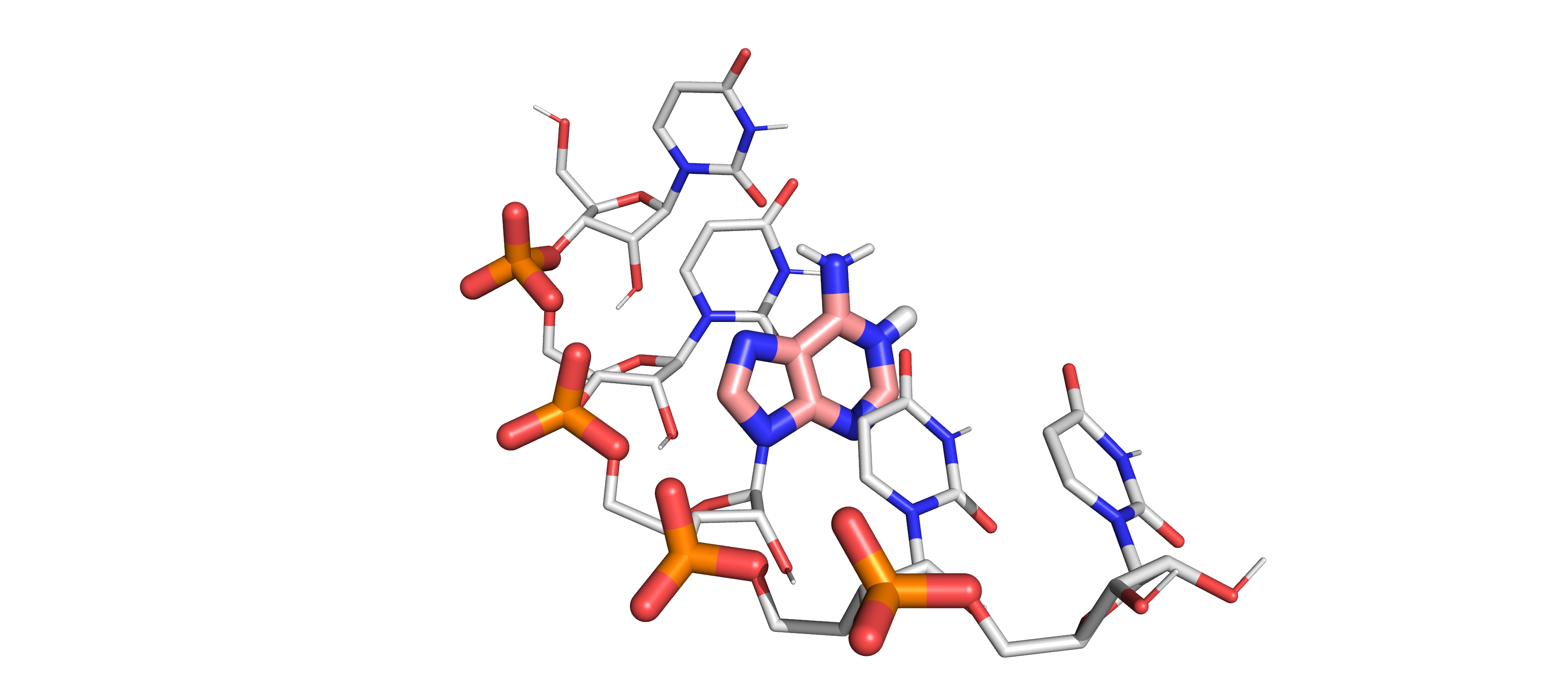}
\end{subfigure}%
\hfill
\begin{subfigure}
    \centering
    \includegraphics[width=0.45\textwidth]{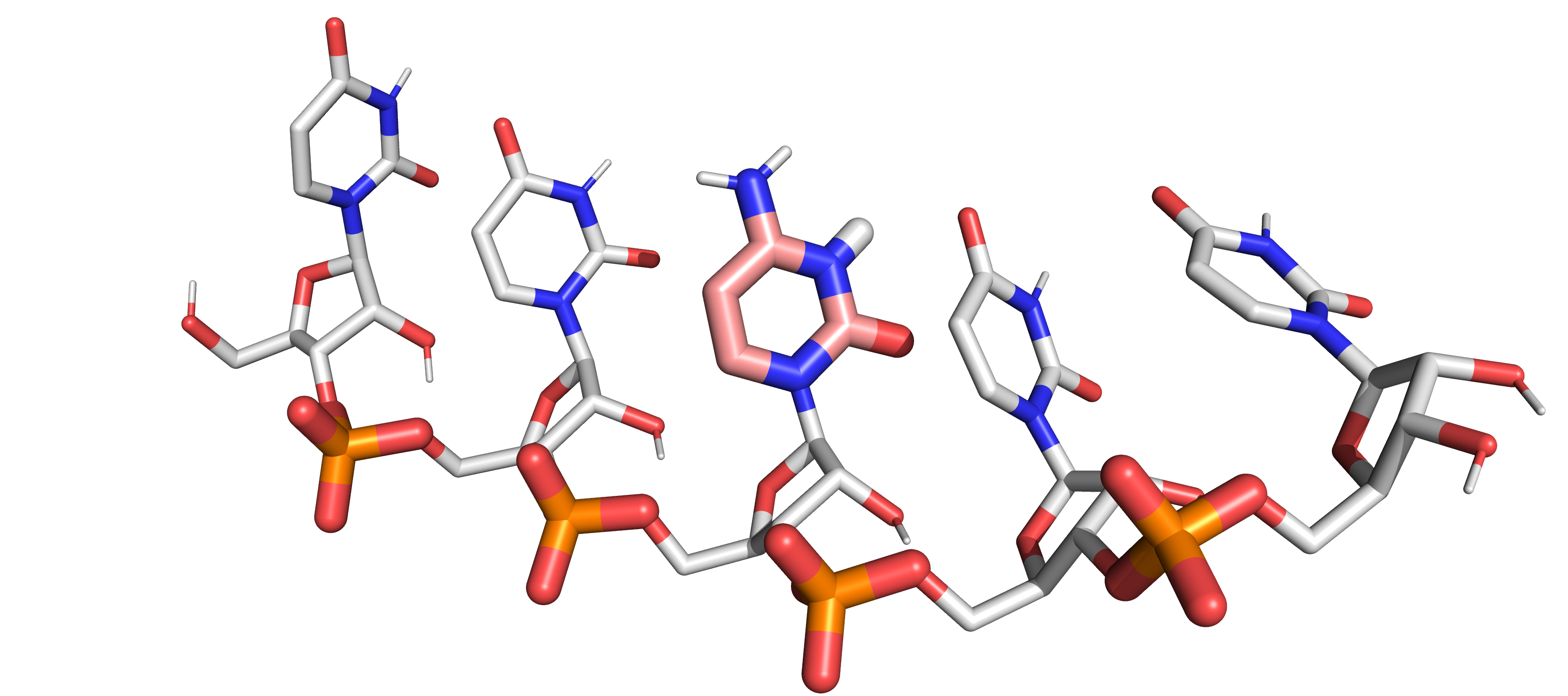}
\end{subfigure}%
\hfill
\begin{subfigure}
    \centering
    \includegraphics[width=0.45\textwidth]{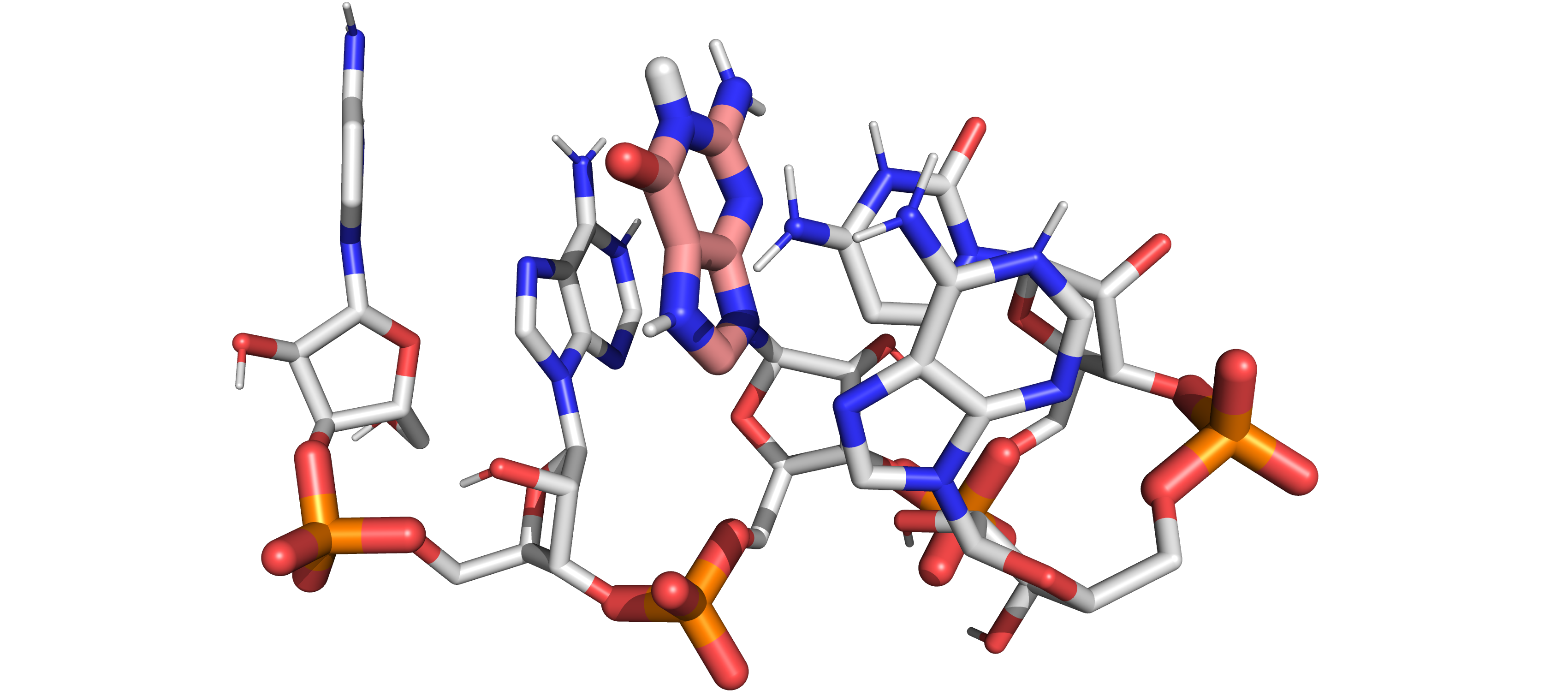}
\hfill
\begin{subfigure}
    \centering
    \includegraphics[width=0.45\textwidth]{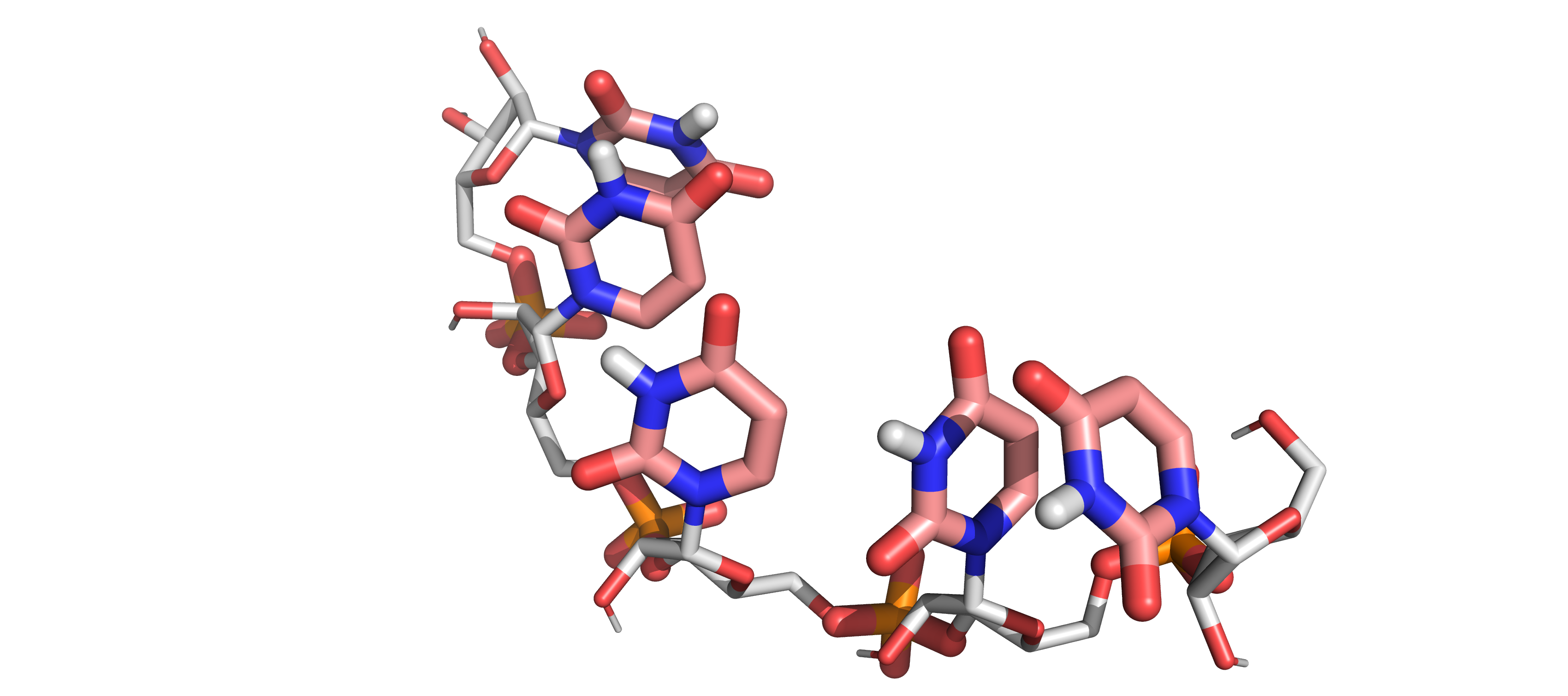}
\end{subfigure}%
\end{subfigure}%
\caption{Cartoon representation of the pentamer systems simulated in this work (see Table~\ref{Table-SI:simsettings}). The titrable nucleobases and proton representations are described in Figure~\ref{Fig:systems1}. Non-polar hydrogens are not represented for improved visual clarity. The selected frames are representative of the energy basin closest to the A-form helix.}
\label{Fig:systems2}
\end{figure}

Test systems were chosen to match available experimental data (Figures~\ref{Fig:systems1} and ~\ref{Fig:systems2}). 
Namely: 
Single-stranded rU\textbf{[A, C]}U and rUU\textbf{[A, C]}UU for the protonable systems were based on the work of Gonzalez-Olvera et al.~\cite{gonzalez2018}. Although this study focused on DNA strands, and absolute \pKa{} values may differ, the relative $\Delta$\pKa{} shifts due to phosphates should be consistent across systems relative to the single nucleoside.
Single-stranded 
rA\textbf{G}C,d rCA\textbf{G}CA, rC\textbf{U}C and 
r\textbf{UUUUU} were constructed to test guanosine~\cite{gonzalez2015,acharya2004} and uridine~\cite{clauwaert1968} deprotonation.

Each system was built using PyMOL~\cite{PyMOL} with the neutral protonation states. All systems were placed in a solvated rhombic dodecahedron box. As highlighted by Sequeira et al.~\cite{sequeira2022}, an important note on adapting force fields such as AMBER or CHARMM to the st-CpHMD is the treatment of long-range electrostatics with PME and the charge variation associated with titration events. Hence, each system net charge was neutralized using the appropriate number of counter-ions for the neutral state of the oligomer and the experimental ionic strength. PME background correction was used to compensate for the charge variations due to these systems' titration.

\subsection{\pKmod{} calibration} \label{sec:pkmod_calib}
The choice of model compound is an important step in the CpHMD method. Protonation and deprotonation events occur on the nucleobase, at N1 or N3, for purines and pyrimidines respectively, with limited long-range electronic effects on the remaining atoms. Hence, the model compound charge variation was restricted to the nucleobase fragment. Under this assumption, our modified force field is constructed within a modular-based rationale, where the phosphate group, the ribose, the hydroxyl group cap (3$^{\prime}$ and 5$^{\prime}$), and each nucleobase are defined as unique residues. The major advantage of this modular approach is that
future inclusion of modified nucleobases in the force field will only require defining the nucleobases and their respective protonation states following this work's protocol.

In this work, we calibrated the model compounds \pK{} (\pKmod{}) applying a similar rationale used in the calibration of titrable amino acid residues~\cite{machuqueiro2011,teixeira2016,sequeira2022} though with an additional initial step. In the first step, we assigned an initial \pKmod{} value to run an iterative procedure of short 3$\times$100~ns CpHMD simulations (adenine) or 1$\times$100~ns CpH-MetaD of all nucleosides at equivalent experimental conditions (300~K and 0.1M). The CVs biased in the metadynamics simulations were the $\chi$ glycosidic angle to promote \textit{syn}/\textit{anti} transitions and a sugar puckering variable to promote transitions between the C2' and C3' endo states~\cite{huang2014}. Single 100~ns CpH-MetaD runs exhibited faster conformational convergence compared to the 3$\times$100~nsCpHMD runs (see Figure~\ref{Fig-SI:CpH_MetaD} in Supporting Information). After obtaining the titration curve, we corrected possible \pKa{} shift relative to the initial guess \pKmod{} after each iteration, until a final \pKmod{} that reproduces the experimental \pKa{} data is obtained. The resulting shifts were smaller than 0.15~pH units, indicating that the electrostatic potentials derived from the PB properly describe the titrable site and the surrounding environment in the nucleobase systems.

The second step occurred after the oligonucleotide CpH-metaD simulations (see Table~\ref{Table-SI:simsettings} for all oligonucleotide sequences and simulation parameters). As referenced in the Introduction section and further discussed in the Results section, the phosphate backbone modulates the protonation behavior of the nucleobases in a more complex biomolecular environment. Similarly to amino acid calibration protocols~\cite{machuqueiro2011}, we applied \textit{a posteriori} corrections to the final \pKmod{} values to accurately reproduce the experimental \pKa{} values and \pKa{} shifts referenced in experimental data.

\subsection{MM/MD and CpHMD Settings}

CpHMD and CpH-MetaD simulations were run using the 2022.3 version of the GROMACS package~\cite{abraham2015,bauer2022} and the open-source, community-developed PLUMED library~\cite{plumed2019} version 2.8.1 version~\cite{tribello2014}. All simulations used the previously described modified $\chi$OL3 AMBER force field~\cite{zgarbova2011} with the OPC water model~\cite{izadi2014}.  

A verlet 1.0~nm cutoff scheme was applied for the PME treatment of non-bonded interactions. Van der Waals interactions were truncated at 10~\AA{}. The integrator time step was 2~fs and the conformations were sampled from an NPT ensemble. Unless otherwise specified, the used temperature bath scheme was the v-rescale~\cite{bussi2007}, at 300~K with a relaxation time of 0.1~ps, coupled to the solute and solvent separately. The system pressure was kept constant with a c-rescale barostat~\cite{bernetti2020} at 1~bar, with a relaxation time of 2~ps and a compressibility of 4.5$\times10^{-5}$~bar$^{-1}$.

\subsection{Poisson-Boltzmann/Monte-Carlo Simulations}

The Delphi V5.1 program\cite{rocchia2002} was used to perform Poisson-Boltzmann calculations. The solute molecular surface was defined by a 1.4~\AA{} radius probe, an ion-exclusion layer of 2.0~\AA{}, and an ionic strength of 0.1~M or 0.01~M depending on the experimental conditions of each system (see Table~\ref{Table-SI:simsettings} in Supplementary Information). The dielectric constants used were 2 and 80, for the solute and solvent, respectively. A two-step focusing procedure was conducted for electrostatic potential calculations by defining two grids of 91~vertices. The coarse grid had a 1~\AA{} spacing between the grid points, while the smaller grid had 0.25 \AA{}. The defined relaxation parameters were 0.20 and 0.75, for linear and non-linear interaction, respectively. Background interaction calculations were not truncated and the electrostatic potential convergence threshold was 0.01~kT/e~\cite{teixeira2014, vilavicosa2015, santos2015}.

The PETIT program performed the MC calculations of the residues' protonation states, using the free energy terms obtained from the PB calculations~\cite{baptista2001}. For each conformation, $10^{5}$~MC cycles were performed and each cycle corresponds to a trial change of each site and pairs of sites with an interaction larger than 2~p\textit{K} units.

\subsection{Metadynamics integration and settings }

In this work, we integrated the well-tempered metadynamics algorithm within the MD production phase of the CpHMD cycle. In well-tempered metadynamics, the system is biased by a smoothly converging history-dependent potential~\cite{barducci2008} along a chosen CV. 

During the MD phase of each cycle, metadynamics was restarted from the bias potential generated in the previous cycle
and deposited new Gaussian potentials. Restarts were performed reading the potential from a grid.
The final conformation is saved to be used in the PB/MC calculation.

In the oligomer simulations, we ran single 1.5 and 3~$\mu$s CpH-MetaD simulations for the trimer and pentamer systems. The CVs biased in the metadynamics simulations were the $\chi$ glycosidic angle and the eRMSD~\cite{bottaro2014,bottaro2016}. The glycosidic torsion promotes transitions between phosphate-exposed (\textit{syn}) and phosphate-shielded (\textit{anti}) states. The eRMSD relates to the relative arrangement between nucleobases in a molecule to a reference fully stacked conformation, thus it is a quantitative measure of base-stacking interactions in single-strand RNA molecules. Each system's pH range and number of simulations were specifically chosen to interpolate their titration curve and \pKa{}. Specific details of each system can be found in the Supporting Information (see Table~\ref{Table-SI:simsettings}).

\subsection{Analyses and error calculations}

\subsubsection{$^{3}$J scalar couplings in nucleosides}

The Karplus equations back-calculate the $^{3}$J scalar couplings from the torsional angles. Experimental reference data~\cite{ancian2010} was used to validate the ensembles for each nucleoside. The parameters for the Karplus equations for $\chi$ and $\chi$' torsional angles were obtained by Munzarova et al.~\cite{munzarova2003}, while the sugar parameters were obtained by Condon et al.~\cite{condon2015}. 

\subsubsection{Structural characterization of the RNA oligonucleotides}

We characterized the oligonucleotides' structure using the chosen collective variables: the eRMSD, a nucleic acid-specific measurement that evaluates nucleobases' orientation and their relative positions, and the glycosidic $\chi$ angle. The analyses were done using the PLUMED software and then reweighted to obtain unbiased populations \cite{branduardi2012}.

\subsubsection{\pKatitle{} calculations and electrostatic contributions}

We estimated the \pKa{} values by obtaining a titration curve for each system and then taking the mid-titration point. The titration curve was obtained by fitting the average protonations of each CpHMD simulation to the Henderson-Hasselbalch (HH) equation~(eq.~\ref{eq:prot}):

\begin{equation}
<P> = \frac{1}{1+10^{pH-\pKa{}}} \text{,}
\label{eq:prot}
\end{equation}
where $<P>$ is the average protonation, pH is the assigned simulation pH and \pKa{} is the fitted parameter.
For the CpH-metaD simulations, the average protonation was
obtained using the reweighting procedure introduced in~\cite{branduardi2012}:

\begin{equation}
<P^*>= \frac{\sum_{i} P_i \times e^{\frac{B_i}{kBT}}}{\sum_i e^{\frac{B_i}{kBT}}}\ \text{,}
\label{eq:weighted_prot}
\end{equation}
 where $<P^*>$ is the weighted average protonation for a given CpH-MetaD simulation and $B_i$ is the
 bias accumulated at the end of the metadynamics simulations calculated on the coordinates corresponding
 to the $i$-th frame.
These weighted average protonations were fitted to the HH equation.

In a different approach, we used binless WHAM~\cite{souaille2001,shirts2008,tan2012} to derive pH-dependent properties such as the \pKa{}, pH scans of the 2D energy maps, and the average protonations of the found energy minima. The procedure consisted of concatenating all CpH-MetaD equilibrated trajectories of a given system and then recomputing the biases for each of the simulations' accumulated bias potentials with the concatenated trajectory. Afterward, we correct each bias with a protonation-dependent contribution for each frame:

\begin{equation}
M_{k,i}^{Bias} = \frac{B_{k,i}}{k_{B}T} + pH_k \times P_{i} \times log(10) \text{,} 
\label{eq:bias_matrix}
\end{equation}
thus obtaining a bias matrix with the biases for each reweighted bias potential ($k$) of the concatenated trajectory dependent on each frame's protonation state ($P_i$) and the simulation pH ($pH_{k}$). Then, we compute the weights using this bias matrix and the protonation states of the concatenated trajectory, using a binless WHAM implementation~\url{https://bussilab.github.io/doc-py-bussilab/bussilab/wham.html}. After using this procedure, we can compute weights at arbitrary pH values to reweight any observable property, such as 2D energy maps, first solvation shell, \pKa{}, and average protonations.

The \pKa{} error values were calculated using a bootstrap approach. For the HH fits, we partitioned each protonation time series into 10 equally sized blocks and then performed bootstrap on the blocks with 500 iterations. We performed a new HH fit for each resample, thus calculating a new \pKa{}. The error was obtained from the standard deviation of the resampled \pKa{} histogram. For the WHAM \pKa{} values, we partitioned the protonations and biases of each pH simulation into 4 blocks and then performed bootstrap (500 iterations). Each resample generated new weights, which were then used to calculate new \pKa{} values. The final error was calculated as previously.

Analysis of time series of observables was performed using the GROMACS tool package. Further analyses were performed using in-house Python scripts or previously specified modules. Block analysis was done for each observable and then the error was obtained through the bootstrap procedure (1000 iterations).

\section{Results}
\label{sec:res}
In this work, we parametrized charges for protonated and deprotonated nucleotides that are relevant in the pH range 3.0--5.0 and 9.0--11.0, respectively, calibrated their \pKmod{} using their respective aqueous \pKa{}{}'s, and tested them in several oligomers for which experimental data is available. One key factor for the nucleobases' \pKa{}'s is the electrostatic effect of the phosphate backbone, which stabilizes 
 positively charged states (adenine, cytidine) and destabilizes negatively charged states (guanosine, uridine). Single-stranded oligonucleotides, possessing multiple phosphate groups, are ideal for validating our method's ability to capture these backbone-dependent protonation variations and to reproduce experimental \pKa{} values accurately. Each titrable nucleotide was simulated within trimer and pentamer systems to capture the phosphate-dependent electrostatics effect on the nucleobase ($\Delta$\pKa{}). Moreover, the chosen flanking residues are non-titrable within the pH range of interest (except in the uridine pentamer system), to prevent protonation coupling effects with other titrable residues. Simulations were done using a combination of constant-pH molecular dynamics and metadynamics. In the following, we report the results of the MD simulations for the adenine and uridine systems. Other systems' results are presented in Supplementary Information.

\subsection{Testing force field modularity}

As mentioned in the Methods section, we redesigned the force field by partitioning each nucleotide chemical moiety (phosphate, sugar, and nucleobase) into a separate residue. This force field modularity aims to facilitate the future inclusion of modified nucleobases. However, the procedure required a small charge rebalance around the C1$^\prime$, H1$^\prime$, and N9/N1 atoms (purines/pyrimidines, respectively). The root-mean-square error of $^3J$ scalar couplings concerning experimental nuclear magnetic resonance data (in Tables~\ref{Table-SI:RMSE_Adenosine}-~\ref{Table-SI:RMSE_Uridine} of Supplementary Information) shows that the modular charge set is comparable to the original force field when assessing the deviation against experiment. Furthermore, the
populations of \textit{syn} and \textit{anti} states of the glycosidic bond angle
are very close to those obtained with the original charge set. These results indicate that enforcing modularity did not affect the overall force-field quality. 

\subsection{Protonable Nucleobases - Adenine \& Cytosine}
\label{sec:prot}

After the calibration steps (see the \pKmod{} calibration section of Methods),
the protonable nucleobases, Adenine (A - \pKa{}~3.5) and Cytosine (C - \pKa{}~4.2) were tested using equivalent trimer and pentamer systems, rU\textbf{A}U/rUU\textbf{A}UU and rU\textbf{C}U/rUU\textbf{C}UU, using our CpH-metaD approach.
\begin{figure}[H]
\centering
\begin{subfigure}
    \centering
    \includegraphics[width=0.45\textwidth]{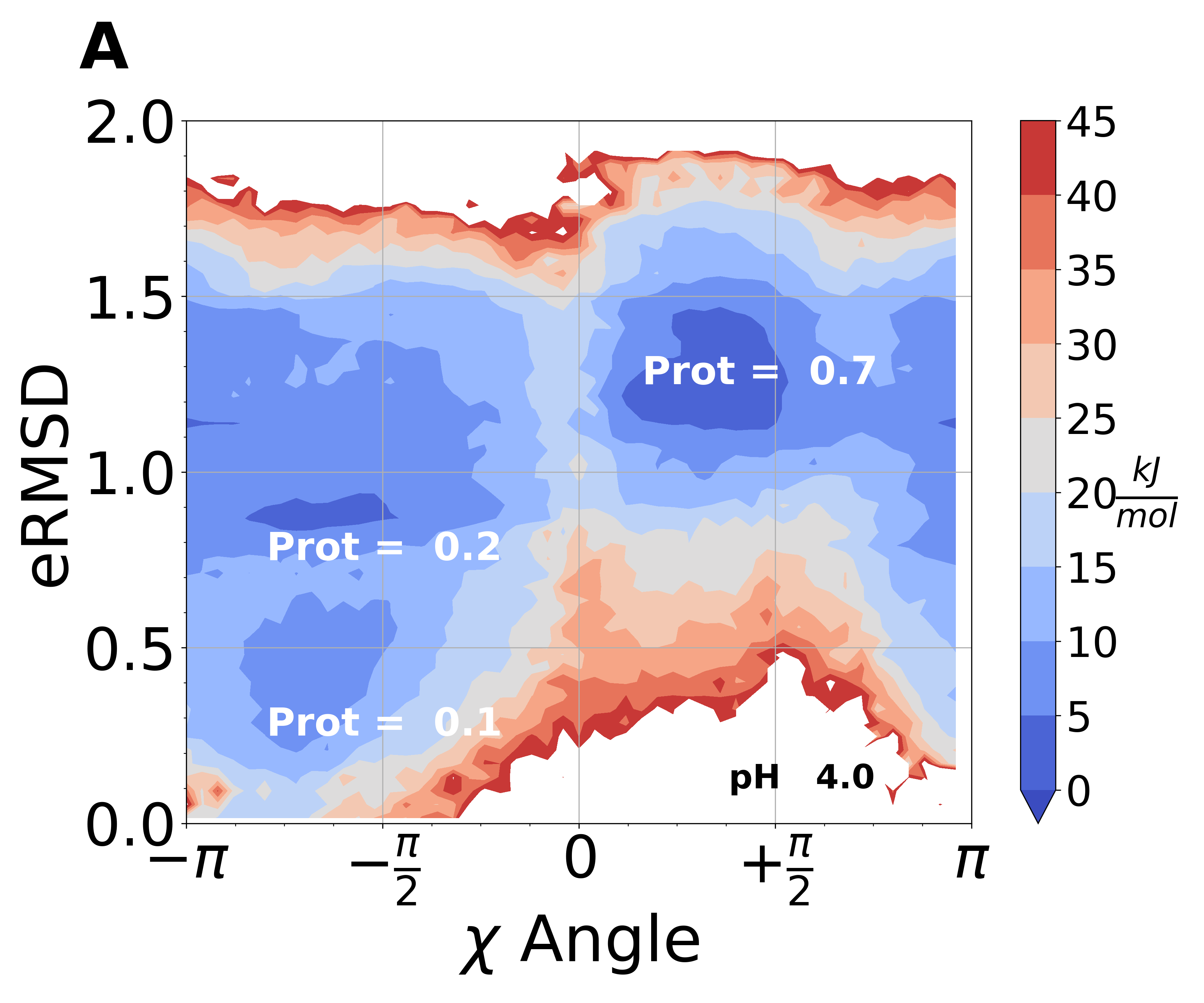}
\end{subfigure}%
\hfill
\begin{subfigure}
    \centering
    \includegraphics[width=0.45\textwidth]{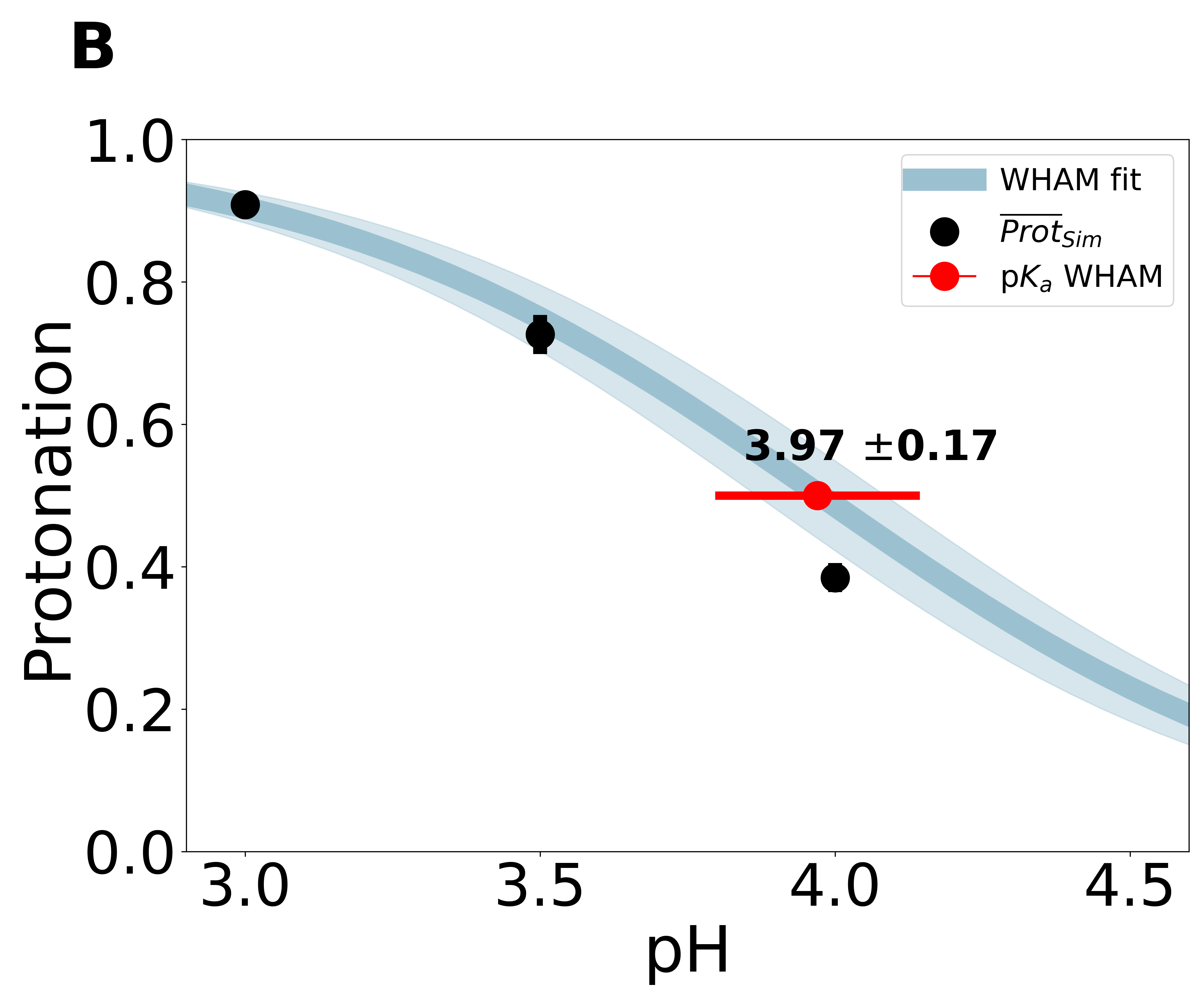}
\end{subfigure}%
\hfill
\begin{subfigure}
    \centering
    \includegraphics[width=0.45\textwidth]{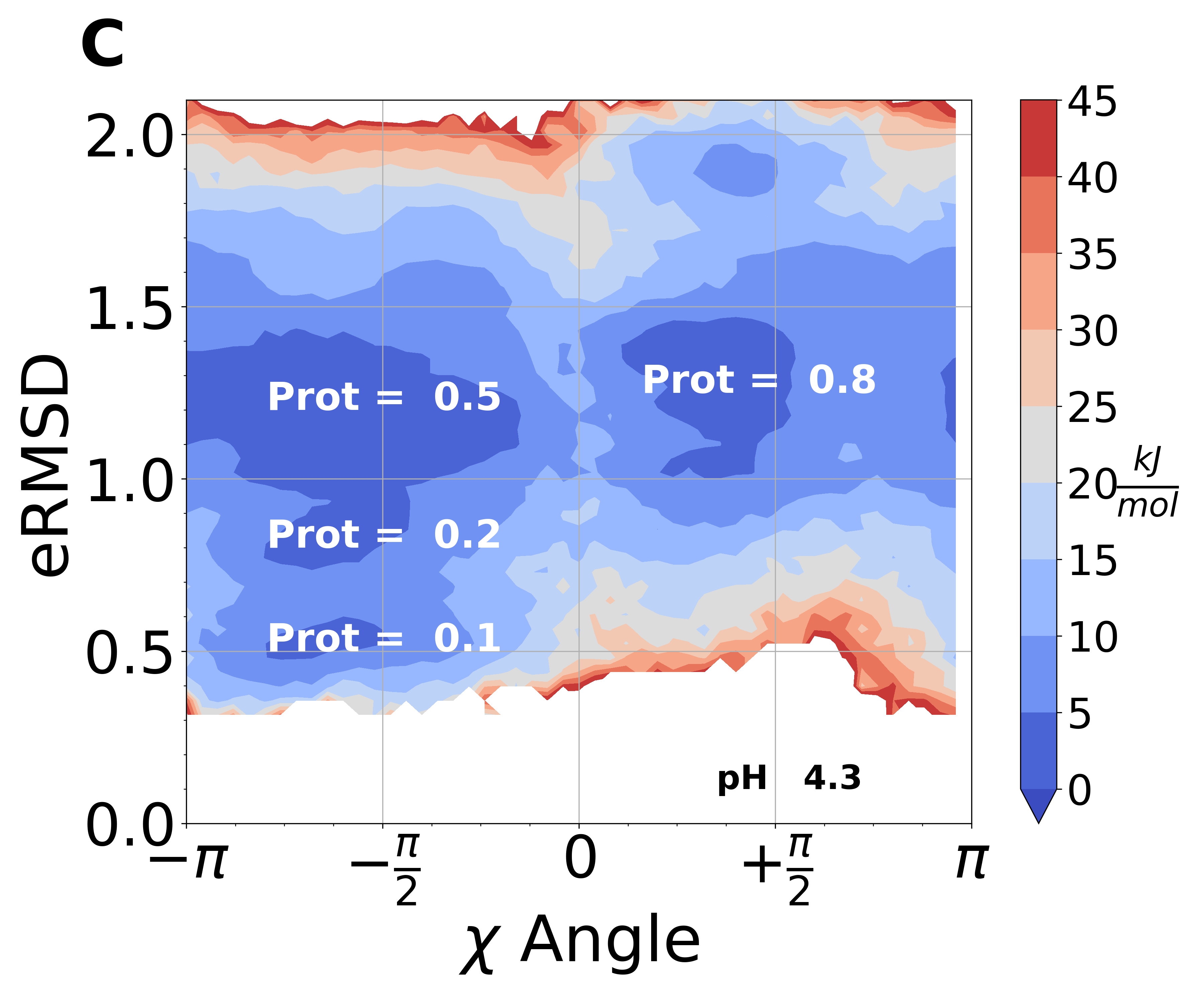}
\hfill
\begin{subfigure}
    \centering
    \includegraphics[width=0.45\textwidth]{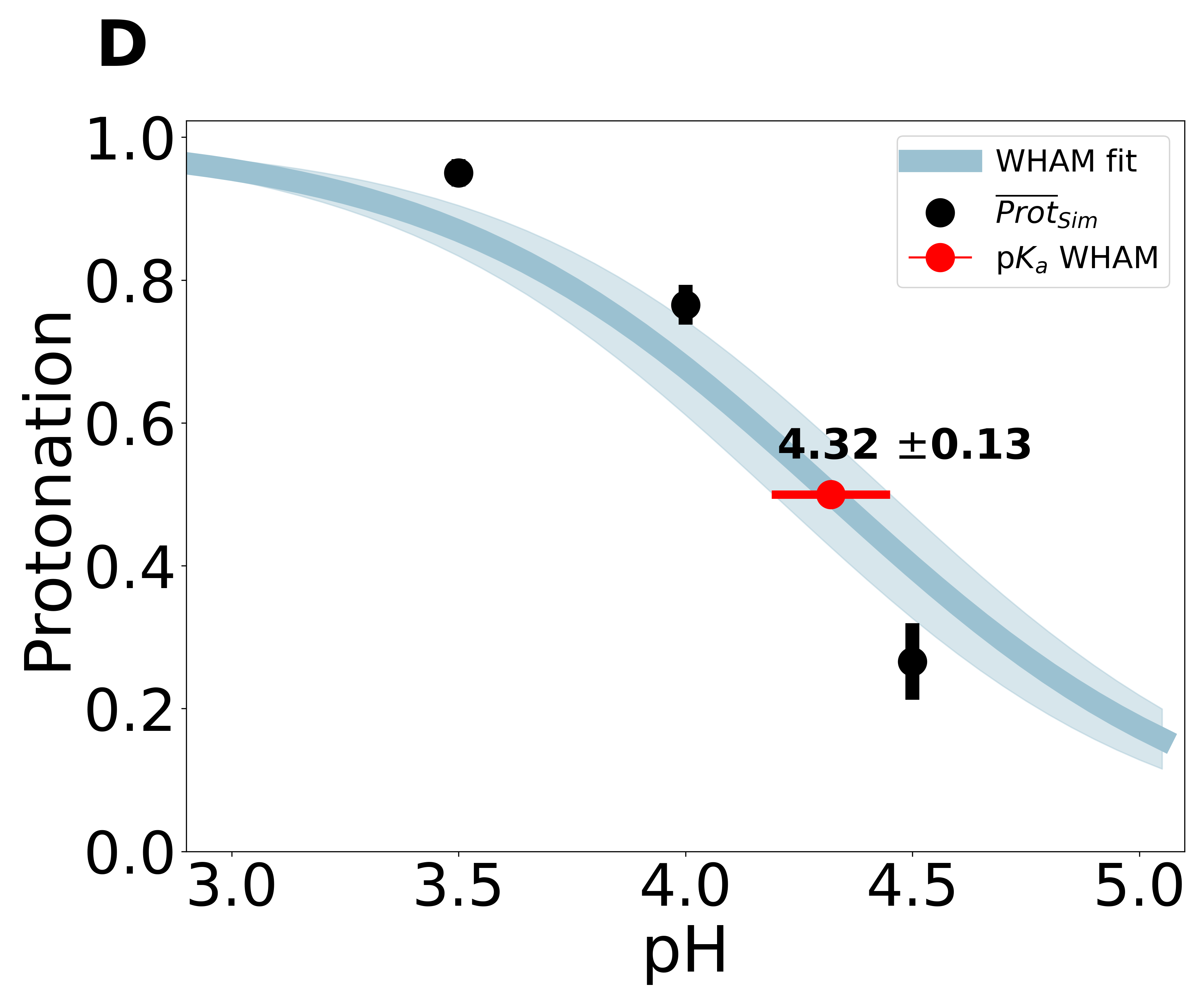}
\end{subfigure}%
\end{subfigure}%
\caption{CpH-MetaD simulations estimate the \pKa{} shifts in the adenine systems. Panels \textbf{A} and \textbf{C} depict 2D free-energy maps as a function of the $\chi{}$ angle of the titrable adenine and the rupture of stacking interactions (eRMSD). The major energy minima are labeled with their average protonation at pH=4.0 and 4.3, respectively. Panels \textbf{B} and \textbf{D}  report the titration curves (gray lines) of the rU\textbf{A}U/rUU\textbf{A}UU as obtained through a binless WHAM reweighting procedure using the equilibrated data from all CpH simulations. The corresponding \pKa{} values are indicated as red circles. Average protonation and respective errors of individual simulations are plotted as black circles. Standard errors were estimated using a bootstrap method.}
\label{fig:Adenine_plots}
\end{figure}
Figure~\ref{fig:Adenine_plots}A reports the
2D free-energy map for the rU\textbf{A}U system at pH~4.0, which is close to the measured \pKa{} value,  as a function of the eRMSD from A-helix, which anti-correlates with nucleobase stacking, and of the $\chi$ angle of the titrable nucleobase. The map displays two distinct energy minima in the \textit{anti} region of the titrable nucleobase at low ($<$0.7) and medium  (0.7$<$ eRMSD $<$ 1.0) eRMSD values, and a single major energy minimum for the \textit{syn} states ($\chi$ angle between $0$ and $+\frac{\pi}{2}$) at high eRMSD values (1.0$<$eRMSD$<$1.5). Visual analysis of the representative structures of each minimum shows progressive unstacking from fully stacked at low eRMSD regions to fully unstacked at high eRMSD regions (Figure~\ref{Fig-SI:A3_structs} of Supplementary Information). This interpretation applies to all systems by observing the representative structures and the 2D maps shown in Supplementary Information.
Average protonations can be seen to be highly conformation-dependent. As stacking interactions are progressively lost (Figure~\ref{Fig-SI:A3_structs}), the average protonation increases, with the highest protonation value usually corresponding to the \textit{syn}/unstacked conformation. Indeed, progressively raising the pH medium leads to more energetically favorable \textit {anti} conformations, while also penalizing the higher protonation \textit{syn} states (Figure~\ref{Fig-SI:A3_2D_maps}). Similar protonation-dependent behavior was observed for the rU\textbf{C}U system as shown in the Supplementary Information (Figures~\ref{Fig-SI:C3_2D_maps}-~\ref{Fig-SI:C3_structs}).
 The rUU\textbf{A}UU 2D free-energy map at pH~4.3, which is close to the measured \pKa{} value, 
 displays 3 minima in the \textit{anti} region of the titrable nucleobase and a single energy minimum for the \textit{syn} state
 Figure~\ref{fig:Adenine_plots}),
Average protonation is anti-correlated with conformational stacking  (Figure~\ref{Fig-SI:A5_structs}), as evidenced by the high average protonation (+0.8) of \textit{syn} unstacked states relative to stacked \textit{anti} states.  Progressively more basic pH environments energetically favor the lower eRMSD states, as seen in Figure~\ref{Fig-SI:A5_2D_maps}. These behaviors are consistent with those observed in the trimer.

Our calculations predict the following \pKa{}'s for the protonable trimers:
3.97 $\pm$ 0.18 for rU\textbf{A}U (Figure~\ref{fig:Adenine_plots}B) and 4.57 $\pm$ 0.13  for rU\textbf{C}U (Figure~\ref{Fig-SI:Cytosine_plots}B), see also table~\ref{Table-SI:allpkas}. Both trimers' \pKa{} similarly deviate from the monomers' values leading to a  $\Delta$\pKa{} (0.35/0.32) that is larger than the experimental $\Delta$\pKa{} (0/0.1 for rU\textbf{A}U/rU\textbf{C}U)~\cite{gonzalez2018}. Interestingly, our model properly captures the electrostatic changes in the titrating residue environment from a trimer to a pentamer. The pentamers' measured \pKa{}'s are 4.32~$\pm$ 0.14 for rUU\textbf{A}UU and 5.07~$\pm$ 0.18 forrUU\textbf{C}UU, see also table~\ref{Table-SI:allpkas}. The resulting $\Delta$\pKa{}'s against the trimers are +0.35 and +0.50, for the rUU\textbf{A}UU  and rUU\textbf{C}UU, respectively. These results are compatible with the $\Delta$\pKa{}$^{exp}$ (A: +0.38 and C: +0.55) for both systems, despite overestimating their absolute \pKa{} values~\cite{gonzalez2018}. For the rUU\textbf{C}UU, we used the HH fit \pKa{} as a reference, due to discrepancies in the WHAM \pKa{} value (see Discussion section).

Since our calculations were not able to reproduce the correct \pKa{} shift between the nucleoside and the trinucleotide, we suggest recalibrating the \pKmod{} using experimental data for trimers (\pKmod{}$_{correction}$). As these trimers are DNA constructs, an additional correction had to be included. In addition, we need to account for the experimental discrepancy reported in the reference work by Gonzalez-Olvera et al. ~\cite{gonzalez2018} against previous literature ($\Delta$\pKa{}$^{Prev-Ref}$). 
Overall, the corrected \pKmod{} was computed as follows: 
\begin{equation} 
\resizebox{1.0\hsize}{!}{$\mathrm{p}K^{\mathrm{mod}}_{correction} = \left( \pKa{}_{DNA}^{Prev} + \Delta\pKa{}^{RNA-DNA}_{Prev} + \Delta\pKa{}^{Prev-Ref} + \Delta\pKa{}_{Trimer-Nuc}^{Ref}\right)  - \pKa{}_{Trimer}^{Model}$}
\label{eq:pkmod}
\end{equation}
Therefore, the adenine \pKmod{} was corrected by $-0.42$ pH units, while the cytosine \pKmod{} was shifted by $-0.2$. These values are defined as $\Delta$\pKa{}'s measured between a hypothetical RNA trimer and our model (see Table~\ref{Table-SI:pkmod}).

Our model can accurately describe the electrostatics and protonation variations between the protonable trimers and pentamers, whereas it failed to estimate the shift from monomer to trimer. Hence, we recommend to use the final recalibrated \pKmod{} for larger constructs. With this parametrization, the model is not expected to reproduce the experimental \pKa{} for a single nucleoside in water.
\subsection{Deprotonable Nucleobases - Guanosine \& Uridine}
\label{sec:deprot}

For the deprotonable nucleobases, uridine (U - \pKa{} 9.2) and guanine (G - \pKa{} 9.2), we applied the same calibration rationale with different sequences: rA\textbf{G}C and rCA\textbf{G}CA for guanine~\cite{gonzalez2015,acharya2004}; rC\textbf{U}C and r\textbf{UUUUU} for uridine~\cite{clauwaert1968}.
\begin{figure}[H]
\centering
\begin{subfigure}
    \centering
    \includegraphics[width=0.45\textwidth]{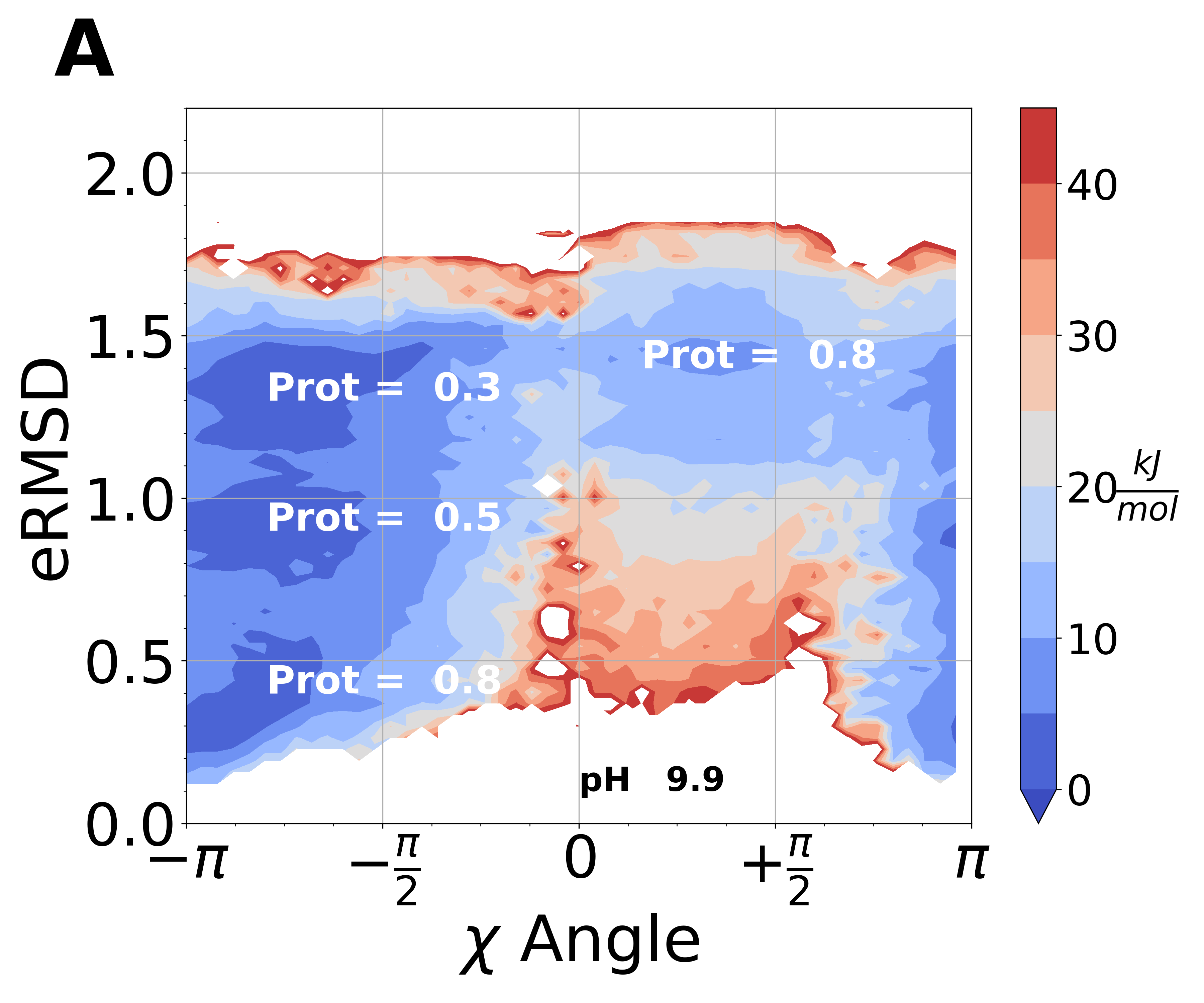}
\end{subfigure}%
\hfill
\begin{subfigure}
    \centering
    \includegraphics[width=0.45\textwidth]{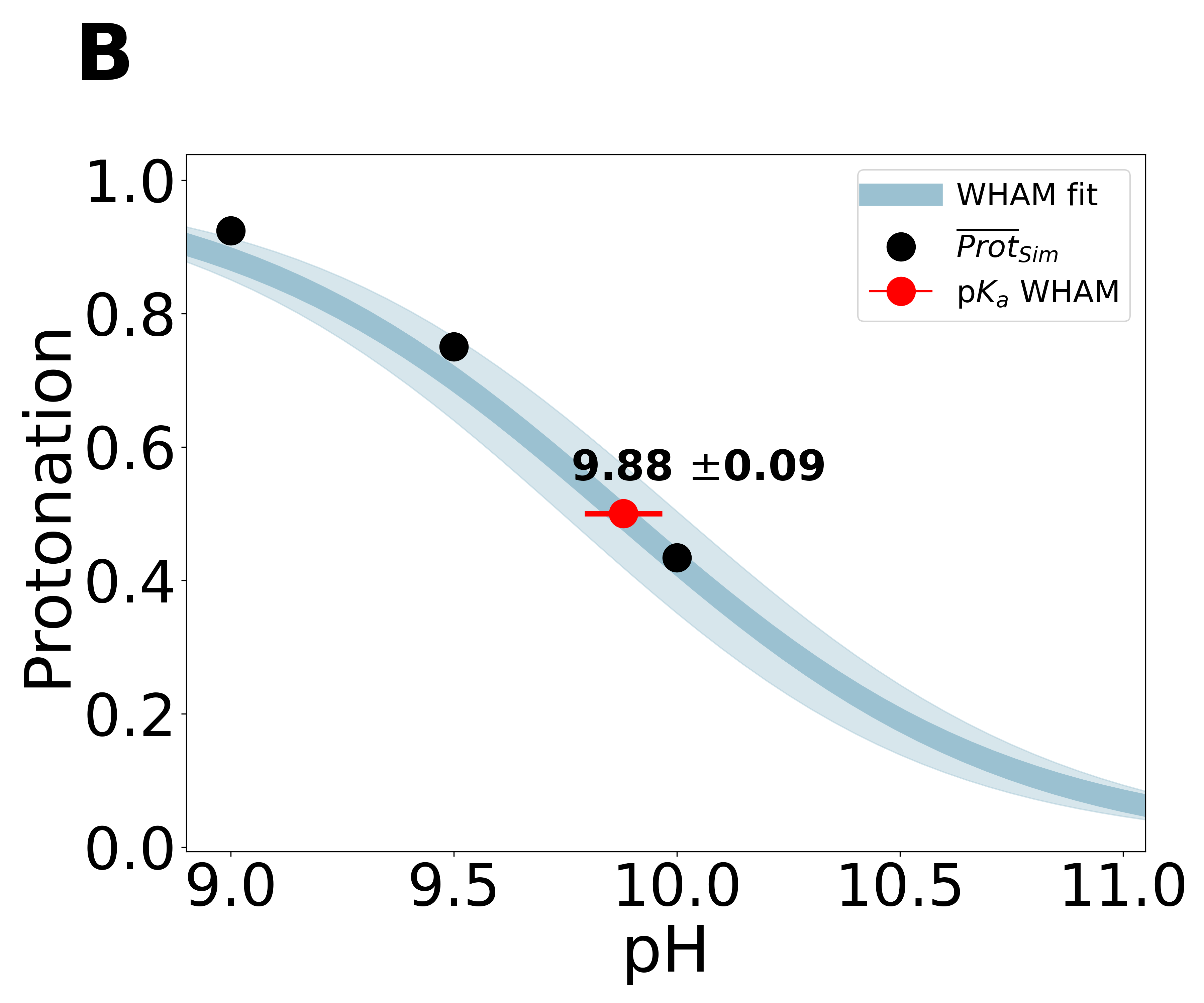}
\end{subfigure}%
\hfill
\begin{subfigure}
 \centering
 \includegraphics[width=0.45\textwidth]
 {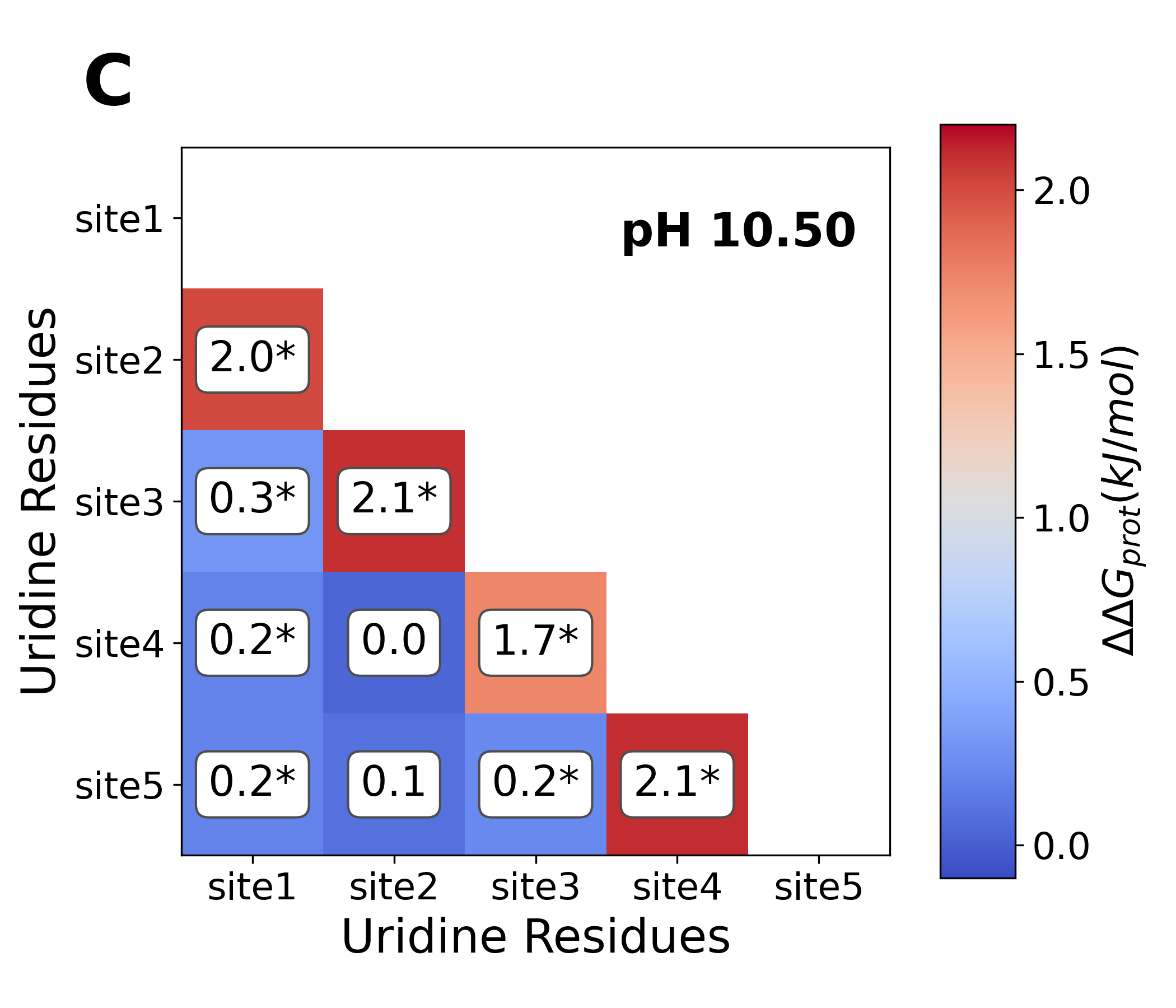}
\end{subfigure}%
\hfill
\begin{subfigure}
    \centering
    \includegraphics[width=0.45\textwidth]{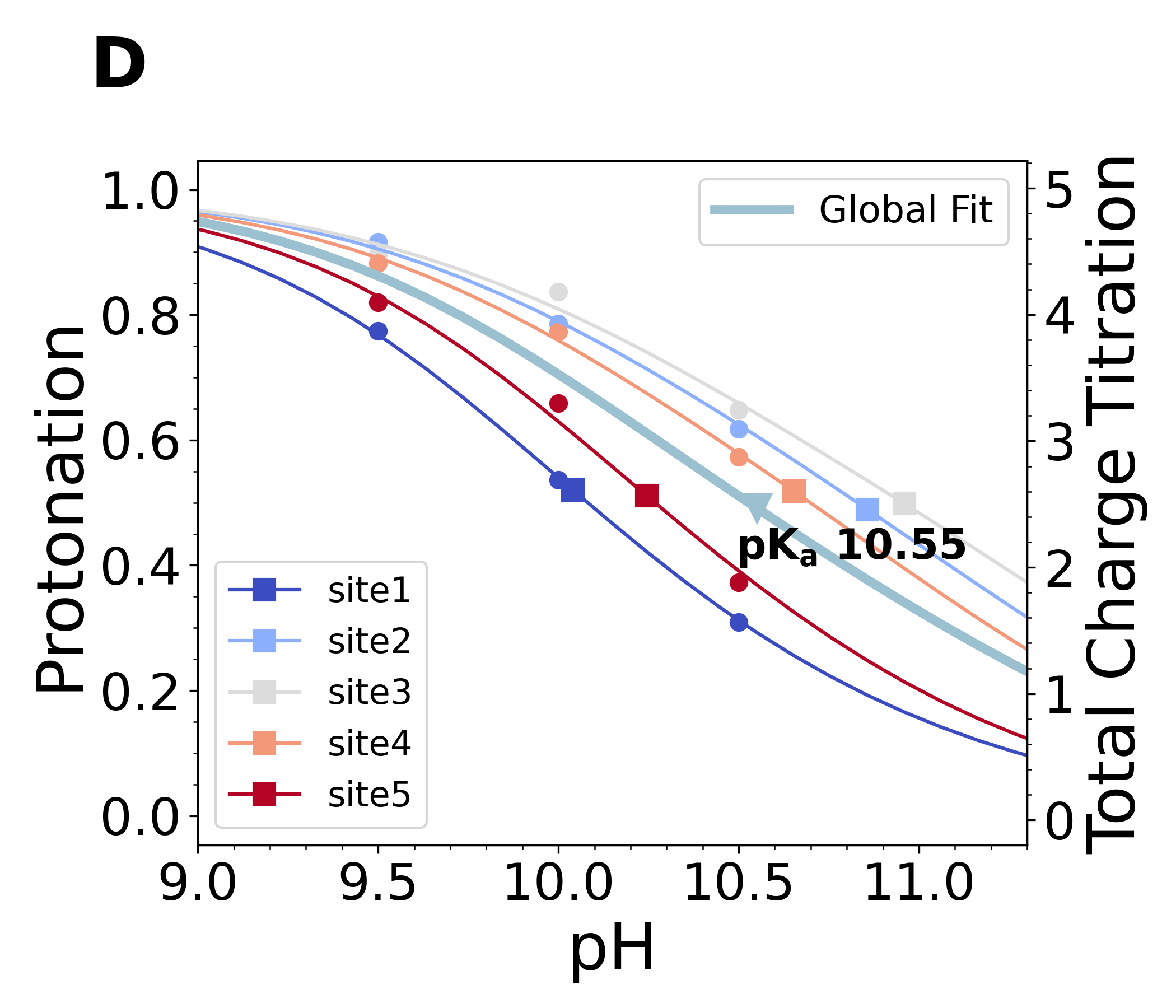}
\end{subfigure}%
\caption{CpH-MetaD simulations estimate the \pKa{} shifts in the uridine systems. Panel A depicts 2D free-energy maps as a function of the rupture of stacking interactions (eRMSD) and the $\chi{}$ angle of the titrable uridine in rC\textbf{U}C.   The major energy minima are labeled with their average protonation at the pH=9.9. Panels B and D report the single and total titration curves (gray lines)  of rC\textbf{U}C and r\textbf{UUUUU}, respectively, as obtained through a binless WHAM reweighting procedure. For rC\textbf{U}C, the \pKa{} value is marked as a red circle, while the individual \pKa{} values of r\textbf{UUUUU}  are labeled as colored squares. These \pKa{} values are compiled in table~\ref{Table-SI:U5mer_pkas} of Supplementary Information. In panel D, the error bars are not displayed for visual clarity. Standard errors were estimated using a bootstrap method. Panel C displays a pairwise protonation cooperativity matrix at pH=10.0 (other matrices in Supplementary Information). The $\Delta\Delta$G indicates the likelihood of simultaneous protonation of two sites in a multi-titrable strand. After a bootstrapping procedure, statistical significance (*) was calculated using the Benjamini-Hochberg method~\cite{benjamini1995}. }
\label{fig:Uridine_plots}
\end{figure}

In the rC\textbf{U}C system, stacking effects correlate with a higher protonation average as shown by the three minima in the \textit{anti} states region ($\chi$ between $-\pi$ and $-\frac{\pi}{2}$). In contrast, unstacked/\textit{syn} states ($\chi$ between $0$ and $+\frac{\pi}{2}$) are, on average, more protonated and less energetically favorable than the other populations. More globular disordered conformations (Figure~\ref{Fig-SI:U3_structs}) tend to expose the titrable group to the solvent and away from the phosphate backbone, facilitating deprotonation events (Figure~\ref{Fig-SI:U3_2D_maps}). 
 Partially stacked conformations (0.8 $<$ eRMSD $<$ 1.0) and fully stacked conformations (Figure~\ref{Fig-SI:U3_structs}) stabilize the protonated state at any pH value as shown by the typically higher average protonations in Figure~\ref{Fig-SI:U3_2D_maps}.   
 Being the titrable site a pyrimidine, \textit{syn} conformations have a very low probability due to steric hindrance.
 This remains true at high pH values.
 The reason why \textit{syn} states are more protonated than their \textit{anti} counterparts is that the titrable N3 faces the negatively charged backbone. Concerning the rA\textbf{G}C system, stacking effects seem more relevant to the thermodynamic stability of the different energy minima than in the other considered systems. Partially stacked (0.8 $<$ eRMSD $<$ 1.0) and alternatively stacked (eRMSD $\approx$ 1.7) conformations are more thermodynamically stable than the trimer's other conformations regardless of the medium pH and average protonation (Figure~\ref{Fig-SI:G3_2D_maps}). Especially, the alternative stacked conformation (Figure~\ref{Fig-SI:G3_structs}) exposes the nucleobase to the solvent, while stacking with the purine ring of the flanking adenosine. These effects strongly stabilize the protonated state in this ensemble, even at high basic media.

While for the previously discussed systems the flanking uridines conferred less stable stacked structures, the rCA\textbf{G}CA showed a smaller likelihood for unstacked conformations. The two identifiable energy minima in the \textit{anti} states region span from the fully to the partially stacked states (Figure~\ref{Fig-SI:G5_structs} and~\ref{Fig-SI:G5_2D_maps} in Supplementary Information). These results hint that partial stacking conformations favor deprotonation events due to a larger solvent exposure than the fully stacked conformations. Meanwhile, in the \textit{syn} region, the proximity to the phosphate backbone increasingly disfavors stacked conformations at more basic mediums (Figure~\ref{Fig-SI:G5_2D_maps} in Supplementary Information). For both the trimer and the pentamer, the thermodynamic stability of rCA\textbf{G}CA conformations with distinct degrees of stacking interactions is less pH sensitive than the previously discussed systems.

In the trimer systems, we obtained slightly deviated estimations from the experimental values for the uridine - \pKa{} 9.88 $\pm$ 0.09 (\pKa{}$^{exp}$ 9.5) - and for the guanosine - \pKa{} 10.25 $\pm$ 0.14 (\pKa{}$^{exp}$ 10.1) (see Table~\ref{Table-SI:allpkas} of Supplementary Information).
Similarly to the protonable nucleobases, the $\Delta$\pKa{} measured between the trimer and the nucleoside - 0.6 and 1.2 - slightly deviated from the experimental $\Delta$\pKa{}$^{exp}$: 0.4~\cite{clauwaert1968} and 0.9~\cite{acharya2004} for rC\textbf{U}C and rA\textbf{G}C, respectively. 

Concerning the rCA\textbf{G}CA system, our \pKa{} prediction of 10.77 $\pm$ 0.23 is close to the experimental \pKa{} of 10.6 (Figure~\ref{Fig-SI:Guanosine_plots}). Similarly to the rUU\textbf{C}UU, the WHAM \pKa{} calculation was significantly underestimated relative to the experimental and the HH fit (Table~\ref{Table-SI:allpkas}), therefore we used the HH fit as a reference.  Still, the $\Delta$\pKa{} between the pentamer and the trimer corroborated well with the $\Delta$\pKa{}$^{exp}$ equal to +0.3 rCA\textbf{G}CA. 

The r\textbf{UUUUU} system presented a distinct challenge due to the increased complexity introduced by simultaneous multi-site titration and site-site coupling effects. The macroscopic \pK{} estimation (10.55 $\pm$ 0.2), extrapolated from fitting the individual charges of the five titrable sites to the Henderson-Hasselbalch equation, was upshifted from the \pKa{}$^{exp}$ (10.0). In this case, we assumed that the midpoint of titration correlated with the macroscopic \pKa{} (Table~\ref{Table-SI:allpkas} of SI) as the experimentally measured signal does not depend on structural effects. Still, the $\Delta$\pKa{} between rC\textbf{U}C and r\textbf{UUUUU} corroborated well with the experimental value (+0.5). 
We also computed each nucleotide's individual \pKa{}s (Table~\ref{Table-SI:U5mer_pkas}) and pairwise cooperativity protonation free energy. These $\Delta{}\Delta{}G$ values indicate the probability of simultaneous protonation events relative to single states titration. Figure~\ref{fig:Uridine_plots}C-D and Figure~\ref{Fig-SI:U5_coop} show that the topological position impacts the simultaneous protonation probability, and consequently the individual \pKa{}s. Terminal residues are more strongly coupled to their neighbor relative to farther nucleotides at all tested pH conditions. The deprotonation of central nucleotides becomes more hindered by the multiple neighbors' titration, hence hindering their deprotonation and reflecting in an increased upshifted \pKa{}, particularly for the third site, as shown for the central uridines in Figure~\ref{fig:Uridine_plots}D.

Similarly to the previous systems, our calculations did not capture the \pKa{} shift from nucleoside to trinucleoside, hence a similar recalibration procedure was applied. However, for the deprotonable systems, the simulated systems are directly comparable to the experimental data. Hence, the \pKmod{} correction shown in equation~\ref{eq:pkmod} is simplified to the $\Delta$\pKa{} between the experimental reference and the model trimer. The guanosine \pKmod{} was corrected by -0.15, while the uridine \pKmod{} was shifted by -0.38 (see Table~\ref{Table-SI:pkmod}).

As in the protonable systems, the method accurately predicts the \pKa{} variations between the trimer and pentamer deprotonable systems, while failing to estimate the change to a nucleoside. Remarkably, the method can estimate individual \pKa{} of coupled titrable sites and the macroscopic midpoint of titration with good agreement with experimental data.

\section{Discussion}
In this work, we developed parameters for protonated and deprotonated nucleotides suitable for constant-pH MD simulations of RNA using the $\chi$OL3 force field, one of the most commonly adopted parametrizations.
Additionally, we integrated well-tempered metadynamics within the st-CpHMD method to facilitate the simultaneous sampling
of conformational and protonation states. Enhanced sampling is crucial for the accurate calibration of nucleic acid parameters. Even short single-strand oligomers require extensive sampling to sufficiently explore their conformational space.
Several factors influence nucleobase titration, such as other titrating chemical moieties, charged neighbors, pH, solvation, and structural effects. These factors alter the free-energy landscape and the likelihood of conformational rearrangements or binding events with other molecules, such as ions, ligands or proteins. The st-CpHMD method balances these diverse contributions, offering an accurate and robust description of structural and protonation dynamics
in a setting that is representative of
the environment in larger RNA molecules. In practice, the pH modulates the sampling probabilities of a given conformation by shifting the thermodynamic equilibrium between energy minima driven by charge variations in the residues. As a result of this protonation-conformation coupling, protonation space sampling can be indirectly improved by enhancing conformational sampling in a given collective-variable (CV) space at a defined pH.  

Our parametrization follows a modular approach compatible with the existing force field, simplifying the future parametrization of modified nucleobases in both the force field and DelPhi databases. Concerning the DelPhi databases parametrization, the reported parameters follow previously established protocols for standard st-CpHMD methodology and can be further optimized if needed in future iterations.
Our nucleobase \pKmod{} calibration protocol focused on these oligomer constructs, where limited conformational sampling could reduce \pKa{} prediction accuracy and hinder the validation against experimental data. The integration of well-tempered metadynamics within the CpHMD method (CpH-MetaD) was pivotal in our calibration procedure, promoting convergence in the simulated RNA oligomers. This enhanced sampling method can be complementary to others such as pH replica exchange~\cite{itoh2011,swails2014,vilavicosa2018}.  %
However, despite this approach, achieving convergence of the rUU\textbf{C}UU and rCA\textbf{G}CA pentamers remained challenging. Consequently, inconsistencies arose in the binless WHAM approach that resulted in underestimated \pKa{} values (Table~\ref{Table-SI:allpkas}). Future work may involve optimizing metadynamics parameters or exploring alternative enhanced sampling techniques to address this issue. Nonetheless, we were able to recover the experimental \pKa{} shifts when analyzing the individual simulations, without assuming overlapping conformational spaces.
The MD simulations performed on a series of test systems allowed the extraction of some general coupling properties between conformation and protonation in RNA oligomers. The \pKa{} of titrable sites is strongly influenced by their proximity to charged phosphate groups. In longer oligomers with high phosphate content, an upward shift in the global \pKa{} consistent with experiments~\cite{acharya2003,gonzalez2015,gonzalez2018} is observed.
Stacking interactions of flanking residues can shield phosphate electrostatics, reducing protonation events. At the same time, at lower pH values, protonation events are favored, leading to increased interactions with neighboring phosphate and reducing the population of stacked conformations. The \textit{syn/anti} transition of the titrable site also impacts the \pKa. In shorter strands, close phosphate contacts of the titrable site are dictated by \textit{syn} states, further upshifting the \pKa{}. Hypothetically, the direction of the shift would be reversed for a titrable site located in a position of the nucleobase that gets farther from the phosphate group upon transitioning to \textit{syn}.
 In the deprotonable systems, the method effectively captures the effect of the flanking residues in strands whose structural dynamics is less sensitive to pH changes, such as rA\textbf{G}C/rCA\textbf{G}CA. In these cases, the energy minima are less affected by variations in the average protonation relative to the stacking effects. Additionally, our approach can also grasp the correlation between multiple deprotonation events, as shown for polyU. Simultaneous titration of multiple nucleotides provides a valuable advantage in dissecting contributions and effects of single sites, a task often difficult for experimental techniques. Our data reveals distinct \pKa{} shifts for uridines at different topological positions, each exposed to slightly distinct electrostatic neighbors. These topological effects have been measured experimentally for other sequences~\cite{acharya2003,acharya2004}, corroborating our results that quantitatively highlight that titration correlation is substantial only between nearest neighbors. Overall, our CpHMD simulations robustly reproduced experimental data and provided novel insight into molecular conformations of RNA single-strands that fall outside the scope of experiments.

Concerning the calibration protocol, two particular notes must be raised for the polyU and the guanosine systems. The experimental data does not specify a pentamer system as representative of polyU. However, the authors discarded the presence of tetra-, tri-, di- and mononucleotides in their polyU experiments. Hence we assumed that a pentamer system was a minimal model to predict the experimental \pKa{} value. Another conflicting point arises from \pKa{} measurements of titrable guanosines: Gonzalez-Olvera et al.~\cite{gonzalez2015}  report \pKa{}~9.2 and 9.7 for trimer and pentamer, respectively. In contrast, Acharya et al.~\cite{acharya2004}  report values of 10.1 and 10.6. Throughout this work, we only considered the NMR measurements of Acharya et al.~\cite{acharya2004} for two reasons: they are directly comparable to our RNA systems and the multiple aromatic markers for \pKa{} measurements confer higher confidence in the reported values.

Importantly, our calculation shows that the adopted CpHMD method does not reproduce the \pKa{} shifts from monomers to trimers, but instead can reproduce correctly the shifts from trimers to pentamers.
For this reason, we decided to recalibrate our \pKmod{} on the trimer simulations, at the expense of having a model that cannot reproduce single nucleosides in water. This correction resembles the one enforced for amino acid \pKmod{} calibration protocols using pentapeptide constructs as shown in \cite{machuqueiro2011}. We consider this a minor defect of the model. These recalibrated \pKmod{} are recommended to be used in further simulation studies.

Our work differs from previous ones on nucleic acid CpHMD, which focused on reproducing the \pKa{} nucleoside in water~\cite{goh2012}.
Compared to Goh et al.~\cite{goh2012}, who focused on \pKa{} shifts in monophosphorylated nucleotides, our study incorporates both backbone and flanking interactions, leading to a more comprehensive description of RNA macromolecular environments. Hence, the accuracy of both procedures is not directly comparable. Finally, the enhanced sampling strategy employed here offers improved convergence of the conformational and protonation landscape, which might be more difficult with pH replica exchange methods alone. Whereas in this work we only tested well-tempered metadynamics, PLUMED integration opens the way to a large class of enhanced sampling methods based on collective variables.

\section{Conclusions}
\label{sec:conc}
Constant-pH molecular dynamics methods are powerful techniques for incorporating titration effects in \textit{in silico} biomolecular studies.  In this work, we successfully extended the st-CpHMD methodology to nucleic acids, achieving accurate predictions of protonation states and \pKa{} values in short oligonucleotides. The seamless integration of well-tempered metadynamics with PLUMED (CpH-MetaD) enhanced convergence and enabled more efficient sampling of protonation-conformation coupling. Usually, neutral conformations are assumed to be the more thermodynamically stable within the physiological pH range. However, these conformations solely represent part of the free-energy landscape. The strength of CpH methods lies in their ability to capture protonation-dependent relative probabilities of conformational states, including higher energy states that might be biologically relevant. These molecular insights are otherwise difficult to obtain without explicitly considering protonation effects at the same time as conformational sampling.

Our results agree with experimental \pKa{} shifts across small oligonucleotides, including complex systems like polyU with multiple coupled titrating sites. Despite initial overestimation of absolute \pKa{} values, \textit{a posteriori} \pKmod{} corrections improved calibration across all nucleobases within biomolecular environments. These results validate both the standard st-CpHMD methodology applied to nucleic acids and the metadynamics integration in the production MD step. These findings present the CpH-metaD approach as a robust tool available for studying pH-dependent processes in nucleic acids, paving the way to study larger, biologically relevant RNA systems, such as ribozymes or RNA-protein complexes, where pH fluctuations play an important role in function and stability. Future work may focus on improving the parameters of metadynamics and Poisson-Boltzmann calculations or applying other enhanced sampling techniques to improve \pKa{} accuracy and conformational or protonation space sampling across more complex systems. 

\section*{Data and Software Availability}
\label{sec:software}
The GROMACS package is freely available software to perform MD simulations and can be downloaded at \url{https://manual.gromacs.org/documentation/2020.1/download.html}. PyMOL~v2.5 is also free software for molecular visualization and generating high-quality images. It can be downloaded from \url{https://pymol.org/2}. The WHAM module is available at ~\url{https://bussilab.github.io/doc-py-bussilab/bussilab/wham.html}. The modified nucleic acid parameters for the st-CpHMD are available at:~\url{https://github.com/Tomfersil/CpH-MetaD}.

All the input scripts are available at: ~\url{https://github.com/Tomfersil/CpH-MetaD}. PLUMED input files are also available on the PLUMED-NEST~\cite{plumed2019} with ID 24.026. A tutorial explaining how to setup and perform analysis of constant-pH metadynamics simulations
is also available on the \url{https://www.plumed-tutorials.org/} website with ID 24.020.

\begin{acknowledgement}
We acknowledge Dr. Miguel Machuqueiro and Nuno F.B. Oliveira for providing the st-CpHMD code, their implementation assistance, and fruitful discussions. We acknowledge financial support from the European Molecular Biology Organization (EMBO) through grant ALTF-399/2022 and the services and resources provided by CINECA through a SISSA-CINECA agreement. 

\end{acknowledgement}

\begin{suppinfo}

\end{suppinfo}

\bibliography{manuscript}

\end{document}